\documentclass[11pt,letterpaper]{article}
\usepackage[letterpaper,margin=1in]{geometry}
\usepackage[all,pdf]{xy}
\usepackage{amsfonts}
\usepackage{lscape}
\usepackage{amssymb}
\usepackage{bbm}
\usepackage{color}
\usepackage{amsfonts,amssymb,mathrsfs,amsmath,amssymb,theorem,float}
\usepackage{graphicx}  

\newtheorem{theorem}{Theorem}[section]

\newtheorem{lemma}[theorem]{Lemma}
\newtheorem{alg}[theorem]{Algorithm}

\newtheorem{defi}[theorem]{Definition}

\def\qed{\hfil {\vrule height5pt width2pt depth2pt}}

\def\qed{\hfil {\vrule height5pt width2pt depth2pt}}
\def\bref#1{(\ref{#1})}

\def\qed{\hfil {\vrule height5pt width2pt depth2pt}}
\def\S{\mathcal{S}}

\def\proof{{\noindent\em Proof.\,\,}}

\def\bref#1{(\ref{#1})}

\def\N{{\mathbb N}}
\def\Z{{\mathbb Z}}

\def\X{{\mathbb{X}}}

\def\F{{\mathbb {F}}}

\def\RB{{\mathcal{R}}}

\def\bref#1{(\ref{#1})}

\def\+{ \oplus}
\def\-{\ominus}
\def\*{\otimes}

\def\deg{\hbox{\rm{deg}}}

\def\modp{{\mathbf{mod}}}

\begin{document}



\title{Sparse Polynomial Interpolation Based on Derivative}
\author{Qiao-Long Huang \\
 Research Center for Mathematics and Interdisciplinary Sciences\\ Shandong University, Qingdao, China\\Email: huangqiaolong@sdu.edu.cn
 }
\date{}

\maketitle

\abstract
\vspace{11pt}
In this paper, we propose two new interpolation algorithms for sparse multivariate polynomials represented by a straight-line program(SLP). Both of our algorithms work over any finite fields $\F_q$ with large characteristic.
The first one is a Monte Carlo randomized algorithm. Its arithmetic complexity is linear  in the number $T$ of non-zero terms of $f$, in the number $n$ of variables. If $q$ is $O((nTD)^{(1)})$, where $D$ is the partial degree bound, then our algorithm has better complexity than other existing algorithms.
The second one is a deterministic algorithm. It has better complexity than existing deterministic algorithms over a field with large characteristic. Its arithmetic complexity is quadratic in $n,T,\log D$, i.e., quadratic in the size of the sparse representation.
And we also show that the complexity of our deterministic algorithm is the same as the one of deterministic zero-testing of Bl\"{a}ser et al. \cite{BHLV09} for the polynomial given by an SLP over finite field (for large characteristic).

\section{Introduction}
We consider the problem of interpolating a sparse multivariate polynomial
$$f=c_1m_1+\cdots+c_tm_t\in \F_q[x_1,\dots,x_n]$$
of degree $d$ with $t$ non-zero terms $c_im_i$, where  $c_1,\dots,c_t$ are coefficients over a finite field $\F_q$ and $m_i,i=1,\dots,t$ are distinct monomials. We assume $f$ is given by a straight-line program and that we know bounds $D> \max_{i=1}^n\deg_{x_i} f$ and $T\geq t$. Denote $\mathbf{char}(\F_q)$ to be the characteristic of $\F_q$.

We summarize our results as follows.

\begin{theorem}\label{the-1}
 Let $f\in\F_q[x_1,\dots,x_n]$, where $\F_q$ is a field. Given any straight-line program $\S_f$ of length $L$ that computes $f$, and bounds $T$ and $D$ for the sparsity and partial degree of $f$.
\begin{itemize}
\item If $\mathbf{char}(\F_q)\geq D$, one can find all coefficients and exponents of  $f$, with probability at least $\frac34$, with a cost of $O^\thicksim(LT\log q+nT\log q+T\log^2 q)$ bit operations.\\
\item If $\mathbf{char}(\F_q)>O^\thicksim(n^2TD)$\footnote{More explicitly, $\mathbf{char}(\F_q)>2npD$, where $p$ is the first $N$-th prime and $N=4\max\{1,\lceil n(T-1)\log_2 D\rceil\}$.}, one can find all coefficients and exponents of  $f$,  with a cost of $O^\thicksim(Ln^2T^2\log^2 D\log q)$ bit operations\\
\end{itemize}
\end{theorem}
%
%
%

\subsection{The Straight-Line Program Model and Interpolation}

The straight-line program is a useful abstraction of a computer program without branches. It is also an important model as a theoretical construct.  Our interpolation algorithms work for $n$-variate sparse polynomials $f\in \F_q[x_1,x_2,\dots,x_n]$ given by a division-free straight-line program $\S_f$ defined as follows.

A division-free Straight-Line Program (SLP) over a ring $\RB$ is a branchless sequence of arithmetic instructions that represents a polynomial function.

\begin{defi} A Straight-Line Program(SLP) over a ring $\RB$ with inputs $x_1, \dots, x_n$ is a sequence of arithmetic instructions $\Gamma=(\Gamma_1,\dots,\Gamma_L)$ of the form
\begin{equation}\label{eq-1}
\Gamma_i =(b_i \leftarrow \alpha \star \beta)
\end{equation}
where $\star\in\{+,-,\times\}$ and $\alpha,\beta\in\{x_1,\dots,x_n\}\bigcup\{b_j | j < i\}\bigcup\RB$. We say $b_L$  is the output for a choice of inputs $x_1,\dots, x_n$.
\end{defi}

$L$ denotes the length of an SLP.  Since an SLP gives us a list of arithmetic instructions, we may choose inputs $x_i$ from $\RB$, homomorphic images of $\RB$, or ring extensions. We say $\Gamma$ computes $f\in\RB[x_1,\dots,x_n]$ if it outputs $b_L =f$ given indeterminate inputs $x_1, \dots, x_n$. We write $\S_f$ to denote an SLP that computes $f$.
In our paper, the ring $\RB$ we used is a finite  field $\F_q$.

\subsection{Previous Work}

The sparse interpolation for multivariate polynomials has received considerable interest.
There are two basic models for this problem:
the polynomial is either given as a straight-line program (SLP) \cite{A16,AGR13,AGR14,AGR16,GS09,GR11,HG19,K85,K88}
or a more general black-box \cite{BT88,HG19b,KY88,M95,Z90}.
As pointed out in \cite{GS09}, for black-box polynomials with degree $d$,
there exist no algorithms yet, which works for arbitrary fields and have complexities polynomial in $\log d$,
while for SLP polynomials, it is possible to give algorithms
with complexities polynomial in $\log d$.

Following \cite{AGR13}, we give a brief introduction to some algorithms specifically intended for straight-line programs.
\vspace{11pt}

$\mathbf{The\ Garg-Schost\ Deterministic\ Algorithm}$. Garg and Schost \cite{GS09} gave a deterministic interpolation algorithm for a univariate  polynomial given by an SLP. Their basic idea consists in evaluating the unknown polynomial $f$ at roots of unity; they constructed these roots by working in extensions of the base ring $\RB$ of the form $\RB[x]/(x^{p_i}-1)$, for suitable values of $p_i$, by recovering $f$ from $f\ \modp\ (x^{p_i}-1)$ for $O(T^2\log D)$ different primes $p_i$.  Their algorithm constructed an integer symmetric polynomial with roots at the exponents of $f$:

$$\chi(y) =\prod_{i=1}^t (y-e_i) \in\Z[y]$$
which is then factored to obtain the exponents $e_i$. Their algorithm first finds a good prime: a prime $p$ for which the terms of $f$ remain distinct when reduced modulo $x^p-1$. The image $f\ \modp\ (x^p-1)$  gives us the values $e_i\ \modp\ p$ and hence $\chi(y)\ \modp\ p$.

Giesbrecht and Roche\cite{GR11} gave a probabilistic method to find the good prime. It improves Garg and Schost's algorithm by a factor $\mathcal{O}(T^2)$, but becomes a Las Vegas algorithm.

\vspace{12pt}
$\mathbf{The\ ``Diversified"\ Interpolation\ Algorithm}$.  Giesbrecht and Roche \cite{GR11} introduced the idea diversification. A polynomial $f$ is said to be diverse if its coefficients $c_i$ are pairwise distinct. The authors show that for appropriate random choices of $\alpha$, $f(\alpha x)$ is diverse with probability at least $\frac{1}{2}$. They then try to interpolate the diversified polynomial $f(\alpha x)$.  As $f(\alpha x)$ is diverse, we can recognize which terms in different good images are images of the same term. Thus, as all the $e_i$ are at most $D$, we can get all the exponents $e_i$ by looking at some $O^\thicksim(\log D)$  images of $f$.

\vspace{12pt}
$\mathbf{Deterministic\ Zero\ Testing}$.
 All the Monte Carlo algorithms can be made Las Vegas (i.e., no possibility of erroneous output, but unbounded worst-case running time) by way of deterministic zero-testing. Given a polynomial $f$ represented by a straight-line program, suppose $f^*$ is the output of a Monte Carlo algorithm that interpolate $f$, the following theorem is used to test whether $f=f^*$.

\begin{theorem}[Bl\"{a}ser et al. (2009); Lemma 13]\label{the-4} Let $\mathcal{R}$ be an integral domain, $f,f^*\in\RB[x]$ and $\#(f-f^*)\leq T,\deg (f-f^*)\leq D$, and suppose $f = f^*\ \modp\ (x^p-1)$ for $T\log D$ primes. Then $f = f^*$.
\end{theorem}

Thus, testing the correctness of the output of a Monte Carlo algorithm requires some $O^\thicksim(T \log D)$ probes of degree at most $O^\thicksim(T \log D)$.  We will show that the complexity of the deterministic zero-testing is the same as the one of deterministic interpolation algorithm presented in this paper if $\RB$ is finite field $\F_q$ with large characteristic, (i.e. our deterministic interpolation algorithm is as easy as the deterministic zero-testing of Bl\"{a}ser et al.).

\vspace{11pt}

$\mathbf{Recursive\ Interpolation}$.
Arnold, Giesbrecht and Roche \cite{AGR13}
gave a  faster recursive algorithm, which is the first time to deduce the complexity about $T$ into linear. The chief novelty behind that algorithm is to use smaller primes $p$ with relaxed requirements. Instead of searching for good primes separating all of the non-zero terms of $f$, they probabilistically search for an ``ok" prime $p$ which separates most of the non-zero terms of $f$. Given this prime $p$ they then construct images of the form $f\ \modp\ (x^{pq_j}-1)$, for a set of coprime moduli ${q_1,\dots,q_k}$ whose product exceeds $D$, in order to build those non-colliding non-zero terms of $f$. The resulting polynomial $f^*$ contains these terms, plus possibly a small number of deceptive terms not occurring in $f$, such that $f-f^*$ now has at most $T/2$ non-zero terms. The algorithm then updates the bound $T\leftarrow T/2$ and recursively interpolates the difference $g = f-f^*$.

\vspace{11pt}

$\mathbf{``Recursive+Diversified"\ Interpolation}$.
In Arnold, Giesbrecht, and Roche \cite{AGR14}, their univariate interpolation algorithm works over finite fields.
By combining the idea of diversification, the complexity becomes better. In their algorithm, they choose a set of	$N\in O^\thicksim(\log D)$ ``ok" primes $p_i\in O^\thicksim(T \log D)$, $1 \leq  i \leq N$. Given these primes $p_i$, they then compute images $f_{ij} = f(\alpha_jx)\ \modp\ (x^{p_i} -1)$ for choices of $\alpha_j$ that will (probably) allow them to identify images of like terms of $f$. This approach relies on a more general notion of diversification and they use information from the images $f_{ij}$ to construct at least half of the terms of $f$. At last, they use the iteratively method to build $f$. The complexity of their algorithm is $O^\thicksim(LT\log^2D(\log D+\log q)\log \frac{1}{\varepsilon})$ bit operations. This cost improves on previous methods by a factor of $\log D$, or $\log q$.

\vspace{11pt}

$\mathbf{``Recursive+Diversified+Substitution"\ Interpolation}$
The previous algorithms for sparse interpolation of straight-line programs are essentially univariate algorithms, but can easily be extended to handle multivariate polynomials by use of the well-known Kronecker substitution. In Arnold, Giesbrecht and Roche \cite{AGR16}, they gave a multivariate interpolation combining the the idea of randomize Kronecker substitutions which achieves similar aims as the Kronecker substitution but with decreased degrees for sparse polynomials.
\subsection{Summary of Results}

In this paper, we propose two  interpolation algorithms for polynomials over a finite field $\F_q$ with large characteristic.
Let $f\in\F_q[x_1,\ldots,x_n]$ be a polynomial given by an SLP of length $L$ with a partial degree bound $D$ and a term bound $T$.

Our first algorithm is a Monte Carlo algorithm. If $\mathbf{char}(\F_q)\geq D$, it finds all coefficients and exponents of $f$, with probability at least $3/4$, with a cost of $O^\thicksim(LT\log q+nT\log q+T\log^2 q)$ bit operations.  Denote $\mathbf{inputsize}$ be the input size and $\mathbf{outputsize}$ be the output size. As we know that the input is the straight-line program and the output is the sparse polynomial, the input size and out size are $O(L\log q)$ and $O(nT\log D+T\log q)$, respectively. So if $q$ is $O((TD)^{(1)})$, our complexity is $O^\thicksim(T\cdot \mathbf{inputsize}+\log q\cdot \mathbf{outputsize})$.

Our second algorithm is a deterministic algorithm. If $\mathbf{char}(\F_q)>O^\thicksim(n^2TD)$, it finds all coefficients and exponents of $f$, with a cost of $O^\thicksim(Ln^2T^2\log^2 D\log q)$ bit operations. Referring to Theorem \ref{the-4}, testing the correctness of the output of an interpolation algorithm requires some $O^\thicksim(T \log D)$ probes of degree at most $O^\thicksim(T \log D)$. So it needs probe $f\ \modp\ (x^p-1)$ for $O^\thicksim(T \log D)$ times and $p$ is $O^\thicksim(T \log D)$. By Lemma \ref{lm-1}, it needs $O^\thicksim(LT^2\log D^2\log q)$ bit operations when the ring $\RB$ is the finite field $\F_q$. Since testing $f=f^*$ is the same as testing $f-f^*=0$, the complexity of our deterministic algorithm is the same as the one of Bl\"{a}ser et al. \cite{BHLV09} deterministic zero-testing. In other word, for a univariate polynomial $f$ given by SLP over a finite field $\F_q$ with large characteristic, deterministically interpolating $f$ is as easy as Bl\"{a}ser et al. \cite{BHLV09} deterministic testing $f=0$.

Table \ref{tab-2} gives a comparison of existing algorithms for sparse interpolation of straight-line programs over the finite field $\F_q$. In the table, for Las Vegas algorithms, we give their average complexity; for Monte Carlo algorithms, we fix the probability of failure $\varepsilon$.

\noindent
\begin{table}[ht]
\footnotesize
\centering
\begin{tabular}{c|c|c}
&Bit&Algorithm\\
&Complexity &type \\ \cline{1-3}
Dense& $LD^n\log q$ & Deterministic\\
Garg $\&$ Schost \cite{GS09}&$Ln^2T^4\log^2D\log q$&Deterministic\\
Randomized G $\&$ S \cite{GR11}&$Ln^2T^3\log^2D\log q$&Las Vegas\\
Giesbrecht $\&$ Roche \cite{GR11}&$Ln^2T^2\log^2 D(n\log D+\log q)$&Las Vegas\\
Arnold, Giesbrecht $\&$ Roche \cite{AGR13}&$Ln^3T\log^3 D\log q$&Monte Carlo\\
Arnold, Giesbrecht $\&$ Roche \cite{AGR14}&$LnT\log^2 D(\log D+\log q)+n^\omega T$&Monte Carlo\\
Arnold, Giesbrecht $\&$ Roche \cite{AGR16}&$Ln\log D(T\log D+n)(\log D + \log q) $&Monte Carlo\\
&$+ n^{\omega-1} T\log D+ n^{\omega}\log D$&\\
%
Huang $\&$ Gao\cite{HG19}&$L n^2T^2\log^2 D\log q+LnT^2\log^3 D\log q$&Deterministic\\
Huang $\&$ Gao\cite{HG19} &$LnT\log^2 D(\log q+\log D)$&Monte Carlo\\
\cline{1-3}
This paper(Th.\ref{the-2})($\mathbf{char}(\F_q)\geq D$)&$L T\log q+T\log^2q+nT\log q$&Monte Carlo\\
This paper(Th.\ref{the-13})($\mathbf{char}(\F_q)\geq O^\thicksim(n^2TD)$)&$L n^2T^2\log^2 D\log q$&Deterministic\\

\end{tabular}
\caption{A ``soft-Oh" comparison  for SLP  polynomials over finite field $\F_q$}\label{tab-2}
\end{table}

From the table, our Monte Carlo algorithm is the first one to separate $n$ and $L$ in complexity. And if $q$ is $O^\thicksim((nTD)^{(1)})$, then it is better than all existing algorithms.  Our deterministic algorithm is better than all existing deterministic algorithms.
The other algorithms works for any finite field, while our algorithms can't work for the finite fields with small characteristic.

\section{The Cost of Probing}
\subsection{Probing points}
Baur-Strassen's \cite{BS83} technique
allows us to evaluate the gradient $(\frac{\partial f}{\partial x_1},\dots,\frac{\partial f}{\partial x_n})$ using $O(L)$ operations
in $\mathcal{R}$. Using $O(L+n)$ operations, this provides an algorithm for the simultaneously evaluation of $f, g_1,\dots, g_n$ with $g_i=x_i(\frac{\partial f}{\partial x_i})$ for $i=1,\dots,n$.
\footnote{ I thank \'{E}ric Schost for pointing out it and I also thank an anonymous referee of my ISSAC paper \cite{H19} for pointing out the work of Baur and Strassen}

We summarize it into a lemma.

\begin{lemma}\label{lm-1}\cite{BS83}
 Assume $f(x)\in\F_q[x]$ is given by an SLP $\S_f$ with length $L$. Let $g_i=x_i(\frac{\partial f}{\partial x_i})$ for $i=1,\dots,n$.  For any point $\overrightarrow{\alpha}=(\alpha_1,\dots,\alpha_n)\in \F^{n}_q$, it costs $O^\thicksim(L+n)$ operations in $\F_q$ to probe $f(\overrightarrow{\alpha})$ and $g_1(\overrightarrow{\alpha}),\dots,g_n(\overrightarrow{\alpha})$ from $\S_f$.
\end{lemma}

\subsection{Probing univariate  polynomials $\mod (x^p-1)$}
In our deterministic algorithms, we need to evaluate polynomial $f\in\F_q[x]$ in an extension ring of $\F_q$. More precisely, we want to evaluate $f$ at $p$th roots of unity for various choices of $p$.
This may be regarded as transforming a straight-line program by substituting operations in $\F_q[x]$ with operations in $\F_q[x]/(x^p-1)$, where each element is represented by a polynomial in $\F_q[x]$ of degree less than $p$. Each evaluation of straight-line program for $f$ in $\F_q[x]/(x^p-1)$ is called a probe of degree $p$.

For $p\in\N_{>0}$, denote
\begin{equation}\label{eq-6}
f_p = f(x)\ \modp\ (x^p-1) \in\F_q[x].
\end{equation}

The following lemma is the complexity of probes of a univariate polynomial given by an SLP.

\begin{lemma}\label{lm-1}\cite{H19}
 Assume $f(x)\in\F_q[x]$ is given by an SLP $\S_f$ with length $L$. Let $p$ be a prime, it costs $O^\thicksim(Lp)$ field operations to probe $f_p$ and $(f')_p$ from $\S_f$.
\end{lemma}

\section{Modified Prony Algorithm Based on Derivative}\label{sec-2}
In this section, we give a Monte Carlo interpolation algorithm for polynomials. The algorithm works as follows. This algorithm is inspired by \cite{H19} and given in \cite{GHS20}.

Assume
\begin{equation}\label{eq-1}
f(\X)=\sum_{k=1}^tc_k\X^{\mathbf{e}_k}\in \F_q[\X]
\end{equation}
where $c_k\in\F^*_q$, $\X=(x_1,\dots,x_n)$, $\mathbf{e}_k=(e_{k,1},\dots,e_{k,n})$ are different vectors and  $\X^{\mathbf{e}_k}=x_1^{e_{k,1}}\cdots x_n^{e_{k,n}}$. Assume $\max_{i=1}^n\deg_{x_i}f< D$ and $\#f\leq T$.
%

\begin{defi}\label{def-2}
A point $\overrightarrow{\alpha}=(\alpha_1,\dots,\alpha_n)\in\F^n_q$  is said a good point of $f$ (as in \bref{eq-1}), iff
$$\overrightarrow{\alpha}^{\mathbf{e}_i}\neq \overrightarrow{\alpha}^{\mathbf{e}_j}, \forall\ i\neq j$$
 \end{defi}

Now we assume the following  two conditions are satisfied:

\begin{itemize}

\item $\mathbf{char}(\F_q)\geq D$.

\item A good point $\overrightarrow{\alpha}\in \F_q^n$ of $f$.

\end{itemize}

Denote $v_k=\overrightarrow{\alpha}^{\mathbf{e}_k},k=1,\dots,t$, then

\begin{equation}
a_j=f(\overrightarrow{\alpha}^{(j)})=\sum_{k=1}^tc_kv^j_k\\
\end{equation}
where $\overrightarrow{\alpha}^{(j)}=(\alpha^j_1,\dots,\alpha^j_n)$.

The term locator polynomial $\Lambda(z)$ is defined as follows.
 \begin{eqnarray}
   \Lambda(z)=\prod_{k=1}^{t} (z-v_k) =z^t+\lambda_{t-1}z^{t-1}+\dots+\lambda_1 z+\lambda_0.
\end{eqnarray}

Let $M_k:=(a_{i+j})_{0\leq i,j<k}$. We have the following property of $M_k$.
\begin{lemma}\cite{KY88}
If $k>t$, $\det M_k=0$; if $k=t$, $\det M_k\neq 0$.
\end{lemma}
%
%

According to Ben-Or and Tiwari's algorithm \cite{BT88}, we have

{\small \begin{eqnarray}\label{eq-2}
&&M_t\cdot
\left(
\begin{array}{c}
\lambda_1\\
\lambda_2\\
\vdots \\
\lambda_{t-1}\\
\end{array}
\right)=\left(
\begin{array}{c}
a_t\\
a_{t+1}\\
\vdots \\
a_{2t-1}\\
\end{array}
\right)
\end{eqnarray}}

We can compute the coefficients $\lambda_i$ from the linear system.
The roots of the polynomial $\Lambda(z)$ give the $v_k$.

By choosing the first $t$ evaluations of $f$, we get the following transposed Vandermonde system for the coefficients of $f$.

{\small \begin{eqnarray}\label{eq-3}
\left(\begin{array}{cccc}
1&1&\cdots&1\\
v_1&v_2&\cdots&v_t\\
\vdots&\vdots&\ddots&\vdots\\
v_1^{t-1}&v_2^{t-1}&\cdots&v_t^{t-1}\\
\end{array}\right)
\left(
\begin{array}{c}
c_1 \\
c_2 \\
\vdots \\
c_t\\
\end{array}
\right)=\left(
\begin{array}{c}
a_0\\
a_1\\
\vdots \\
a_{t-1}\\
\end{array}
\right)
\end{eqnarray}}

Now fix $i$, we show how to compute the $i$-th degree of each term in $f$. Consider the evaluations $h_{i,j}=g_i(\overrightarrow{\alpha}^{(j)})=(x_i\frac{\partial f}{\partial x_i})(\overrightarrow{\alpha}^{(j)}),j=0,1,\dots,t-1$.

Now we have the following key theorem.

\begin{theorem}\cite{GHS20}
For $i=1,\dots,n,j=0,1,\dots,t-1$, $$h_{i,j}=\sum_{k=1}^t c_ke_{k,i}v_k^j$$
\end{theorem}

\vspace{11pt}

From the above theorem, we have
{\small \begin{eqnarray}\label{eq-4}
\left(\begin{array}{cccc}
1&1&\cdots&1\\
v_1&v_2&\cdots&v_t\\
\vdots&\vdots&\ddots&\vdots\\
v_1^{t-1}&v_2^{t-1}&\cdots&v_t^{t-1}\\
\end{array}\right)
\left(
\begin{array}{c}
c_1e_{1,i} \\
c_2e_{2,i} \\
\vdots \\
c_te_{t,i}\\
\end{array}
\right)=\left(
\begin{array}{c}
h_{i,0}\\
h_{i,1}\\
\vdots \\
h_{i,t-1}\\
\end{array}
\right)
\end{eqnarray}}

We can compute $c_k,c_ke_{k,i}$ from the above systems \bref{eq-3} and \bref{eq-4} and then compute all $e_{k,i},k=1,\dots,t$.

\subsection{Algorithm}

\begin{alg}[Interpolation Based on Derivative]\label{alg-2}
\end{alg}

{\noindent\bf Input:}
\begin{itemize}
\item An  SLP $\S_f$ that computes $f\in \F_q[\X]$.
\item A sparsity bound $T$ of $f$.
\item A degree bound $D>\max_{i=1}^n\deg_{x_i}\ f$ and $D\leq \mathbf{char}(\F_q)$.
\end{itemize}

{\noindent\bf Output:} Return correct $f=\sum_{k=1}^tc_k\X^{\mathbf{e}_k}$ with probability $\geq\frac34$ or Failure.
\begin{description}
\item[Step 1:] If $q\geq 2DT(T-1)+1$, randomly choose $\overrightarrow{\alpha}$ from $\F^{*n}_{q}$.

If $q<2DT(T-1)+1$, extend $\F_q$ into $\F_{q'}$, where $q'=q^u$ and $u=\lceil\frac{\log_2(2DT^2)}{\log_2 q}\rceil$. Randomly choose $\overrightarrow{\alpha}$ from $\F^{*n}_{q'}$.

\item[Step 2:] For $j=0, \dots, 2T-1$, get the evaluations $a_j=f(\overrightarrow{\alpha}^{(j)})$.

\item[Step 3:]
For $i=1,\dots,n,j=0,1,\dots,T-1$, get the evaluations $h_{i,j}=(x_i\frac{\partial f}{\partial x_i})(\overrightarrow{\alpha}^{(j)})$.

\item[Step 4:]  Find the rank $t$ of the matrix $M_T$ and solve equation \bref{eq-2}  to get the coefficients of the term locator polynomial $\Lambda(z)$.

\item[Step 5:]  Find all  roots  $v_k,k=1,\dots,t$ of $\Lambda(z)$.

\item[Step 6:] Find the coefficients $c_i$ by solving the transposed Vandermonde system \bref{eq-3}.

\item[Step 7:] For $i=1,\dots,n$, solve the transposed Vandermonde system \bref{eq-4} to find the coefficients $e_{k,i}c_k,k=1,\dots,t$.

\item[Step 8:] For $i=1,\dots,n,k=1,\dots,t$, compute $e_{k,i}$ from the division of $c_k$ and $c_ke_{k,i}$.

\item[Step 9:] Return $\sum_{k=1}^tc_kz^{e_{k,1}}_1\cdots z^{e_{k,n}}_n$.

 \end{description}

%
%
%
%

\subsection{Complexity}
Now we analyse the complexity. Since we compute $e_{k,i}$ from the division of $c_k,c_ke_{k,i}$, not from $v_k=\overrightarrow{\alpha}^{\mathbf{e}_k}$, we avoid computing the discrete logarithms over finite field $\F_q$. So we have the following theorem. We assume that we may obtain
a random bit with bit-cost $O(1)$.

\begin{theorem}\label{the-2}
The expected bit complexity of Algorithm \ref{alg-2} is $O^\thicksim(LT\log q+T\log^2 q+nT\log q)$.
\end{theorem}
\proof
In Step 1, if $q\geq 2DT(T-1)+1$, then randomly choose $\overrightarrow{\alpha}$ from $\F^{*n}_q$ and  it costs $O^\thicksim(n\log q)$ bit operations.

If $q<DT(T-1)+1$, suppose the element of $\F_{q^u}$ is represented as $\F_q[x]/\langle \Phi(x)\rangle$, where $\Phi$ is a degree-$u$ irreducible polynomial over $\F_q$. In \cite{S94}, Shoup proved that finding an irreducible polynomial $\Phi$ with degree $u$ over $\F_q$ costs expected $O^\thicksim(u^2\log q+u\log^2 q)$ bit operations. Since $u$ is $O(\frac{\log  T+\log D}{\log q})$, the bit complexity is $O^\thicksim(\frac{\log^2(DT)}{\log q}+\log (DT)\log q)$. Randomly choose $\overrightarrow{\alpha}$ from $\F^{*n}_{q'}$ and  it costs $O^\thicksim(n\log q')=O^\thicksim(nu\log q)$ bit operations.

In the following steps, if $q\geq 2DT(T-1)+1$, the field we used is $\F_q$; otherwise, the field used is $\F_{q'}$. We first consider the case $q\geq 2DT(T-1)+1$.

In Step 2 and Step 3, for a fixed $j$, by Lemma \ref{lm-1}, it needs $O(L+n)$ operations in $\F_q$ to probe $f(\overrightarrow{\alpha}^{(j)})$ and $(x_i\frac{\partial f}{\partial x_i})(\overrightarrow{\alpha}^{(j)}),i=1,\dots,n$. Since $j=0,1,\dots,2T-1$, the total cost of probes is $O(LT+nT)$ operations in $\F_q$ .

In Step 4, since Equ.\bref{eq-2} is a Hankel system, it can be solve by $O^\thicksim(T)$ operations in $\F_q$ \cite{KY88}.

In Step 5, we can compute all $v_1,\dots,v_t$ with expected $O^\thicksim(T\log q)$ operations in $\F_q$. \cite[Cor.14.16]{GG99}

In Step 6, solving the Vandermonde system  costs $O^\thicksim(t)$ operations in $\F_q$.\cite{KY88}

In Step 7, solving the Vandermonde system  costs $O^\thicksim(nt)$ operations in $\F_q$.\cite{KY88}

In Step 8, We compute $e_{k,i}$ from the division of $c_k,c_ke_{k,i}$, so it costs $O^\thicksim(nt)$ operations  in $\F_q$ .

In total, it costs  $O^\thicksim(LT+T\log q+nT)$  operations  in $\F_q$, which  is $O^\thicksim(LT\log q+T\log^2 q+nT\log q)$ bit operations.

In the case of $q< 2DT(T-1)+1$. The field used is $\F_{q'}$, then the total complexity is $O^\thicksim(LT\log q'+T\log^2 q'+nT\log q')$ bit operations. $\log q'=u\log q$ is $O(\log (DT^2))$. Since $\mathbf{char}(\F_q)\geq D$, $\log q\geq \log D$. So $\log q'$ is $O(\log T+\log q)$.
The bit complexity is $O^\thicksim(LT\log q+T\log^2 q+nT\log q)$ bit operations.
\qed

\vspace{11pt}
The following lemma is used to prove the correctness of Algorithm \ref{alg-2}.
\begin{lemma}\label{lm-8}\cite{GHS20}
The probability that a randomly chosen $\overrightarrow{\alpha}\in\F^{*n}_q$ is a good point of $f$ is
$$\geq 1- \frac{DT(T-1)}{2(q-1)}$$
\end{lemma}

\begin{theorem}
Algorithm \ref{alg-2} is correct.
\end{theorem}
\proof
As mentioned before, if $\overrightarrow{\alpha}$ is a good point of $f$, then Algorithm \ref{alg-2} will return the correct polynomial.

As stated in Lemma \ref{lm-8}, we can randomly choose a point from $\F^{*n}_q$. But to ensure a success rate of $\geq \frac34$, we will ensure $q\geq 2DT(T-1)+1$. If $q<2DT(T-1)+1$, we can extend $\F_q$ into the extended field $\F_{q^u}$, where
\begin{equation}
u=\lceil\frac{\log_2(2DT^2)}{\log_2 q}\rceil
\end{equation}

So we have $q'\geq 2DT(T-1)+1$. In Step 1, by Lemma \ref{lm-8}, $\overrightarrow{\alpha}\in \F^{*n}_{q'}$ is a good point with probability $$\geq 1- \frac{DT(T-1)}{2(q-1)}\geq \frac34$$

The correctness is proved.
\qed

\section{Deterministic Interpolation Algorithm}
In this section, we give a deterministic interpolation algorithm for a polynomial given by SLP over  finite field $\F_q$(for large characteristic). The algorithm works as follows.
First, we give a deterministic interpolation algorithm for univariate polynomials.
Then, we use the sparse deterministic Kronecker substitutions to extend the univariate algorithm into a multivariate one.

\subsection{Preliminaries}
Let
$ f=a_1x^{d_1}+\cdots+a_t x^{d_t}\in \F_q[x]$
be a univariate polynomial with non-zero terms $a_ix^{d_i}$. %
Denote $\#f=t$ to be the number of terms of $f$ and $M_f=\{a_1x^{d_1},a_2x^{d_2},\dots,a_tx^{d_t}\}$ be the set of terms in $f$.
Let  $D,T\in\N$ such that $D> \max_{i=1}^n\deg_{x_i}(f)$ and $ T\ge\#f$.

We have the following key concept.
\begin{defi}\label{def-c1}
A term $a_ix^{d_i}\in M_f$  is called a collision in $f_p$ if there exists
an $a_jx^{d_j}\in M_f,j\neq i$ such that $d_i\ \modp\ p=d_j\ \modp\ p$.
\end{defi}

We have the following lemmas.
\begin{lemma}\label{lm-3}\cite{H19}
Let $f\in \F_q[x]$, $\mathbf{char}(\F_q)>\deg f$. Assume $d>0$ and $ax^d\in M_f$. If $ax^d$ is not a collision in $f_p$, then $(ad)x^{d-1}\in M_{f'}$ and $(ad)x^{d-1}$ is not a collision in $(f')_p$.
\end{lemma}


We will give an algorithm to recover the non-collision terms of $f$ from $f_p$ and $(f')_p$.

Denote
\begin{eqnarray}\label{eq-ufdp}
f_p&=&c_0+c_1x^{e_1}+c_2x^{e_2}+\cdots+c_k x^{e_k}\cr
(f')_p&=&h_0x^{p-1}+h_1x^{e_1-1}+h_2x^{e_2-1}+\cdots+h_kx^{e_k-1}+g
\end{eqnarray}
where $0<e_1<e_2< \cdots < e_k$,  $c_i,h_i\in\F_q,i=0,\dots,k$, and $c_i\neq 0,i=1,\dots,k$. The polynomial $g$ contains all the terms in $(f')_p$ whose degrees are not $p-1$ or $e_i-1,i=1,2,\dots,k$.

As in \cite{H19}, we now introduce the following key notation.
\begin{eqnarray}
&&U_{f_p}:=
\{c_i x^{d_i}| \hbox{ such that for some } i\in[0,k] \cr\nonumber
 &&\quad\hbox{U1}: d_i=h_i/c_i \hbox{ and } d_i\in\N.\\ \nonumber
 &&\quad\hbox{U2}:d_i\in[0,D).\}\nonumber
\end{eqnarray}

The following lemma gives the geometric meaning of $U_{f_p}$.
\begin{lemma}\label{lm-4}\cite{H19}
Let $f\in \F_q[x]$, $\mathbf{char}(\F_q)>\deg f$ and $ax^d\in M_f$.
If $ax^d$ is not a collision in $f_p$, then $ax^d\in U_{f_p}$.
\end{lemma}

The following algorithm computes the set $U_{f_p}$.
\begin{alg}[UTerms]\label{alg-1}
\end{alg}

{\noindent\bf Input:}
\begin{itemize}
\item Univariate polynomials $f_p,(f')_p$.

\item A prime $p$.

\item A degree bound $D> \deg (f)$, where $\mathbf{char}(\F_q)\geq D$.
\end{itemize}
{\noindent\bf Output:} $U_{f_p}$.

\begin{lemma}\label{lm-5}\cite{H19}
Algorithm \ref{alg-1} needs $O(T)$ field operations and $O^\thicksim(T\log D)$ bit operations, where $T\geq \#f$.
\end{lemma}

\subsubsection{Recover multivariate non-colliding terms}
Assume $f(x_1,\dots,x_n)\in\F_q[x_1,\dots,x_n]$. Let $p$ be a prime.
Denote
\begin{equation}\label{eq-5}
f(x^\mathbf{s})=f(x^{s_1},x^{s_2},\dots,x^{s_n})
\end{equation}
$$f(x^{\mathbf{s}+p\mathbf{I}_k})=f(x^{s_1},\dots,x^{s_k+p},\dots,x^{s_n})$$ to be the univariate polynomials after substitutions $x_i=x^{s_i},i=1,2,\dots,n$ and substitutions $x_i=x^{s_i},i=1,2,\dots,n,i\neq k,x_k=x^{s_k+p}$, where $\mathbf{I}_k\in\Z_{\geq0}^n$ is the $k$-th unit vector.

\begin{defi} Let $f\in\F_q[x_1,x_2,\dots,x_n]$, $\mathbf{s}=(s_1,s_2,\dots,s_n)\in\N^n$ and prime $p$. $f(x^{\mathbf{s}})$ and $f(x^{\mathbf{s}})_p$ are defined in \bref{eq-5} and \bref{eq-6}.
A term $cm_1$ of $f$ is said to {\em collide} in $f(x^{\mathbf{s}})$
(or $f(x^{\mathbf{s}})_p$) if $f$ has another term $em_2$
such that $m_1\ne m_2$ and $m_1(x^\mathbf{s})=m_2(x^\mathbf{s})$ (or $m_1(x^\mathbf{s})_p=m_2(x^\mathbf{s})_p$).
\end{defi}

Now we describe how to recover the non-collision terms from $f(x^{\mathbf{s}})$ and $f(x^{\mathbf{s}_+p\mathbf{I}_k}),k=1,2,\dots,n$.

%
Let  \begin{eqnarray}\label{eq-t6}
 &&f{(x^\mathbf{s})}\ \modp\ (x^p-1)=a_1x^{d_1}+\cdots+a_r x^{d_r} \label{eq-mfdp}
\end{eqnarray}
Since $f{(x^\mathbf{s})}\ \modp\ (x^p-1) = f{(x^{\mathbf{s}+p\mathbf{I}_k})}\ \modp\ (x^p-1)$, for $k=1,2,\dots,n$,   we can write
  \begin{eqnarray}\label{eq-t7}
 &&f{(x^\mathbf{s})}=f_1+f_2+\cdots+f_r+g\label{eq-mfp}\\
 &&f{(x^{\mathbf{s}+p\mathbf{I}_k})}=f_{k,1}+f_{k,2}+\cdots+f_{k,r}+g_k\nonumber
\end{eqnarray}
where
$f_i\ \modp\ (x^p-1)=f_{k,i}\ \modp\ (x^p-1)=a_ix^{d_i}$, $g\ \modp \ (x^p-1)=g_k\ \modp \ (x^p-1)=0$.
We define the following key notation
\begin{eqnarray}
&&\hbox{TS}_{(f,p,\mathbf{s})} =\{a_i x_1^{e_{i,1}}\cdots x_n^{e_{i,n}}| a_i \hbox{ is from } \bref{eq-mfdp}, \hbox{ and }  \cr
 &&\quad \hbox{T1}: f_i=a_ix^{u_i},f_{k,i}=a_ix^{v_{k,i}},k=1,2,\dots,n.\label{eq-uf2}\\ 
 &&\quad \hbox{T2}: e_{i,k}=\frac{v_{k,i}-u_i}{p}\in\N,k= 1,2,\dots,n.\cr
 &&\quad \hbox{T3}: u_i =e_{i,1}s_1+e_{i,2}s_2+\cdots+e_{i,n}s_n
 .\cr
 &&\quad \hbox{T4}:  \sum_{j=1}^n e_{i,j}\leq D. \}\nonumber
\end{eqnarray}

%

\begin{lemma}\label{lm-5}\cite{H19}
Let $f=\sum_{i=1}^tc_im_i\in\F_q[x_1,x_2,\dots,x_n]$ and $D\geq\deg(f)$. If $cm$ does not collide in $f(x^{\mathbf{s}})_p$, then $cm\in \hbox{TS}_{(f,p,\mathbf{s})}$.
\end{lemma}

\vspace{11pt}

We give the following algorithm to compute $\hbox{TS}_{(f,p,\mathbf{s})}$.
\begin{alg}[TSTerms]\label{alg-4}
\end{alg}

{\noindent\bf Input:}
\begin{itemize}
\item Univariate polynomials $f(x^\mathbf{s})_p,f(x^\mathbf{s}),f(x^{\mathbf{s}+p\mathbf{I}_k})\in\F_q[x]$, where $k=1,2,\dots,n$.

\item A prime $p$.

\item A vector $\mathbf{s}=(s_1,s_2,\dots,s_n)\in \Z^n_{\geq 0}$.

\item Degree bound $D\geq \deg(f)$.
\end{itemize}
{\noindent\bf Output:} $\hbox{TS}_{(f,p,\mathbf{s})}$.

\begin{lemma}\label{lm-6}\cite{HG19b}
Algorithm \ref{alg-4} needs $O(nT)$ arithmetic operations in $\F_q$ and $O^\thicksim(nT\log(s_{\max} D+pD))$ bit operations, where $s_{\max}=\max\{s_1,s_2,\dots,s_n\}$.
\end{lemma}

We will give the reduction algorithm which is used to obtain the polynomials $g(x^{\mathbf{s}+p\mathbf{I}_k}),k=1,\dots,n$ from the exact form of $g(x_1,\dots,x_n)$.

\begin{alg}[PolySubs]\label{alg-5}
\end{alg}

{\noindent\bf Input:}
\begin{itemize}
\item A polynomial  $g\in\F_q[x_1,\dots,x_n]$.

\item A vector $\mathbf{s}=(s_1,\dots,s_n)\in \Z^n_{\geq 0}$.

\item A prime $p$.
\end{itemize}
{\noindent\bf Output:} $g(x^{\mathbf{s}+p\mathbf{I}_k}),k=1,\dots,n$.

\begin{lemma}\label{lm-7}\cite{HG19b}
The complexity of Algorithm \ref{alg-5}  is $O^\thicksim(nt\log (p+s_{\max})+nt\log(\deg (f)))$ bit operations and at most $O(nt)$ arithmetic operations in $\F_q$, where $s_{\max}=\max\{s_1,s_2,\dots,s_n\}$.
\end{lemma}

The following theorem is the key for both the univariate and multivariate interpolation algorithms.

\begin{theorem}\label{the-10}\cite{HG19}
Let $f=\sum_{i=1}^tc_im_i\in\mathcal{F}[x_1,x_2,\dots,x_n]$, $\mathcal{F}$ be a field, $T\geq \# f,D>\max_{i=1}^n\deg_{x_i} f,\mathbf{s}=(1,D,\dots,D^{n-1})$, $N_1=\max\{1,\lceil n(T-1)\log D\rceil\}$, and $p_1,p_2,\dots,p_{4N_1}$ be $4N_1$ different primes.
Let $j_0$ be an integer in $[1,4N_1]$ such that $\# f(x^{\mathbf{s}})_{p_{j_0}}\geq \# f(x^\mathbf{s})_{p_{j}}$ for all $j$.
%
Then $f$ has at least $\lceil\frac t 2\rceil$ terms which are not a collision in $f(x^\mathbf{s})_{p_{j_0}}$.
\end{theorem}

The following theorem gives a technique to determine whether a term belongs to $f$.

\begin{theorem}\label{the-11}\cite{HG19}
Let $f=\sum_{i=1}^t c_im_i\in \mathcal{F}[x_1,x_2,\dots,x_n]$, $\mathcal{F}$ is a field, $T\geq \# f,D>\max_{i=1}^n\deg_{x_i} f,\mathbf{s}=(1,D,\dots,D^{n-1})$, $N_1=\max\{1,\lceil n(T-1)\log D\rceil\}$, $N_2=\lceil nT\log D\rceil$, and $p_1,p_2,\dots,p_{N_1+N_2-1}$ be $N_1+N_2-1$ different primes.
For a term $cm$ satisfying $\deg(m)<D$, $cm\in M_f$ if and only if there exist at least $N_2$
integers $j\in[1,N_1+N_2-1]$ such that $\#(f-cm)(x^\mathbf{s})_{p_j}<\#f(x^\mathbf{s})_{p_j}$.
\end{theorem}

\subsection{The interpolation algorithm for univariate polynomials}
Now we give a deterministic interpolation algorithms for univariate polynomials. The basic idea is:
First, we use Theorem \ref{the-10} to find a prime $p$
such that at least half part of the
terms of $f$ are not collisions in $f_p$.
Then, we use $f_p,(f')_p$
to find a set of terms containing these non-collision terms by the coefficients division and add all these terms into a polynomial $f^*$.
Finally,  recursively interpolates the difference $f-f^*$.

\vspace{11pt}
The following algorithm is used to interpolate at least half number of terms of $f-f^*$.

\begin{alg}[DUIHalf]\label{alg-8}
\end{alg}

{\noindent\bf Input:}
\begin{itemize}
\item An  SLP $\S_f$ that computes $f(x)$, where $f(x)\in \F_q[x]$ and $\mathbf{char}(\F_q)\geq D$.

\item An approximation $f^*$.

\item A terms bound $T\geq \max(\#f,\#f^*)$.

\item A terms bound $T_1\geq \# (f-f^*)$, where $T\geq T_1$.

\item A degree bound $D>\max(\deg_{x_i} f,\deg_{x_i} f^*)$.
\end{itemize}
{\noindent\bf Output:}  $f^{**}$, which satisfies $\#(f-f^*-f^{**})\leq \lfloor\frac{T}{2}\rfloor$ and $\max_{i=1}^n\deg_{x_i} f^{**}<D$.

\begin{description}

\item[Step 1:] Let $N_1=\max\{1,\lceil (T-1)\log_2 D\rceil\},N_2=\lceil T\log_2 D\rceil,N=\max\{4N_1,N_1+N_2-1\}$.

\item[Step 2:] Find the first $N$ primes $p_1,p_2,\dots,p_{N}$.

\item[Step 3:] For $j=1,2,\dots,N$, probe  $f_{p_j}$ from $\S_f$. Let $f_j=f_{p_j}-f^*_{p_j}$.

\item[Step 4:] Let $\alpha=\max\{\#f_j|j=1,2,\dots,N\}$ and $j_0$ be the smallest number such that $\#f_{j_0}=\alpha$.
    If $\alpha=0$, then return $0$; end if;

\item[Step 5:] Probe $(f')_{p_{j_0}}$ from $\S_{f'}$. Let $g=(f')_{p_{j_0}}-(f^{*'})_{p_{j_0}}$.

\item[Step 6:] Let $U_{(f-f^*)_{p_{j_0}}}:=\mathbf{UTerms}(f_{j_0},g,p_{j_0},D)$.

\item[Step 7:] Let $f^{**}:=0$. For each $u\in U_{(f-f^*)_{p_{j_0}}}$,
if
$$\#\{j\,|\, \#(f_j-u_{p_j})<\#(f_j),j=1,\dots,N_1+N_2-1 \}\ge N_2$$
then $f^{**}:=f^{**} + u$.

\item[Step 8:] Return $f^{**}$.

\end{description}

\begin{lemma}\label{lm-9}
Algorithm \ref{alg-8} returns the correct $f^{**}$ using   $O^\thicksim(LT^2\log^2 D)$ field operations in $\F_q$ plus a similar number of bit operations.
\end{lemma}
\proof
By Theorem \ref{the-10}  and Lemma \ref{lm-4},  in Step 6, at least half of the terms of $f-f^*$ are in $U_{(f-f^*)_{p_{j_0}}}$.
In Step 7, Theorem \ref{the-11} is used to select the elements of $M_{f-f^*}$ from $U_{(f-f^*)_{p_{j_0}}}$.
Then, the correctness of the algorithm is proved.

Now we analyse the complexity.

In Step 2, since the bit complexity of finding the first $N$ primes is $O(N\log^2N\log\log N)$ by \cite[p.500,Thm.18.10]{GG99} and $N$ is $O^\thicksim(nT\log D)$, the bit complexity of Step 2 is $O^\thicksim(nT\log D)$.

In Step 3, we probes $N$ times. Since $p_{j}$ is $O^\thicksim(T\log D)$, by Lemma \ref{lm-1}, the cost of the probes is $O^\thicksim(LNT\log D)$ field operations. Since $N$ is $O(T\log D)$, the cost is $O^\thicksim(LT^2\log^2 D)$ field operations.
Since $\#f^*\leq T$, it needs $O^\thicksim(T^2\log D)$ field operations and $O^\thicksim(T^2\log^2 D)$ bit operations to obtain all $f^*_{p_{j}}$ and $f_j$.

In Step 4, we find the integer $j_0$. Since $\#(f-f^*)_{p_{j}}\leq T$, it needs at most $O^\thicksim(NT)$ field operations and $O^\thicksim(NT\log D)$  bit operations to compute all $\#(f-f^*)_{p_{j}},i=1,2,\dots,N$, which is $O^\thicksim(T^2\log D)$ field operations and $O^\thicksim(T^2\log^2 D)$ bit operations.
Find $j_0$ needs $O^\thicksim(T\log D)$ bit operations. So the bit complexity of Step 4 is $O^\thicksim(T^2\log D)$.

In Step 5, by Lemma \ref{lm-1}, it needs $O^\thicksim(LT\log D)$ field operations.

In Step 6, by Lemma \ref{lm-5}, the complexity is $O(T)$ field operations and $O^\thicksim(T\log D)$ bit operations.

In Step 7, in order to determine whether $\#(f_j-u_{p_j})<\#(f_j)$, we just need to determine whether $u_{p_j}$ is a term of $f_j$. We sort the terms of $f_j$ such that they are in ascending order according to their degrees, which costs $O^\thicksim((N_1 + N_2)T\log D)=O^\thicksim(T^2\log^2 D)$ bit operations, since $N_1 +N_2$ is $O(T \log D)$. To find whether $f_j$ has a term with degree $\deg(u_{p_j})$, we need $O(\log T)$ comparisons. Since the height of the degree is $O(\log D)$, it needs $O(\log T\log D)$ bit operations. To compare the coefficients, it needs one arithmetic operation. So it totally needs $O(\log T \log D)$ bit operations and $O(1)$ arithmetic operation to compare $\#(f_j-u_{p_j})$ with $\#(f_j)$. Hence, the total complexity of Step 7 is $O^\thicksim(\#U_{{(f-f^*)}_{p_{j_0}}} (N_1+N_2)\log T \log D + (N_1 + N_2)T\log D)$ bit operations
and $O(\#U_{{(f-f^*)}_{p_{j_0}}} (N_1+N_2))$ field operations. Since $\#U_{{(f-f^*)}_{p_{j_0}}}\leq T$ and $N_1+N_2-1$ is of $O(T \log D)$, the total complexity is $O^\thicksim(T^2 \log D)$ field operations and $O^\thicksim(T^2 \log^2 D)$ bit operations.

We proved the theorem.
\qed
\vspace{11pt}

Now we give the completed univariate interpolation algorithm.

\begin{alg}[DUIPoly]\label{alg-9}
\end{alg}

{\noindent\bf Input:}
\begin{itemize}
\item
SLP $\S_f$ that computes $f(x)\in \F_q[x]$.

\item A terms bound $T\geq \#f,$.

\item A degree bound $D> \max_{i=1}^n\deg_{x_i} f$.
\end{itemize}
{\noindent\bf Output:} The exact form of $f$.

\begin{description}

\item[Step 1:] Let $h=0, T_1=T$.

\item[Step 2:] While $T_1>0$ do

\begin{description}
\item[a:] Let $g=\mathbf{DUIHalf}(\S_f,h,T,T_1,D)$.

\item[b:] Let $h=h+g$, $T_1=\lfloor\frac{T_1}{2}\rfloor$.

\end{description}

\item[Step 3:] Return $h$.

\end{description}

\begin{theorem}\label{the-12}
Algorithm \ref{alg-9} can be used to find $f$ using $O^\thicksim(LT^2\log^2 D)$ field operations plus a similar number of field operations.
\end{theorem}
\proof
Since Algorithm $\mathbf{DUIHalf}$ at least obtain half of $f-h$, it takes $O(\log T)$ times of Step 2. In $\mathbf{a}$, by Lemma \ref{lm-9}, the complexity is $O^\thicksim(LT^2\log^2 D)$ field operations plus a similar number of bit operations.
In $\mathbf{b}$, it needs $O(T)$ field operations and  $O^\thicksim(T\log D)$ bit operations to update $h$. We proved the theorem.
\qed

\subsection{Deterministic multivariate polynomial interpolation}
In this section, assume $f\in\F_q[x_1,\dots,x_n]$ and $D>\max_{i=1}^n\deg_{x_i} f$. To interpolate an $n$-variate polynomial $f(x_1,\dots,x_n)$, we can directly apply a Kronecker substitution, and interpolate
$\widehat{f}(x)=f(x,x^D,\dots,x^{D^{n-1}})$. While this certainly increases the degree, $f$ and $\widehat{f}$ have the same number of nonzero terms, and $f$ can be easily recovered from $\widehat{f}$. This reduces the problem of interpolating the $n$-variate polynomial $f$ of  degree no more $D$ to interpolating a univariate polynomial $\widehat{f}$ of degree at most $D^n$.

But in our univariate interpolation algorithm over finite field $\F_q$, the polynomial to be interpolated needs to satisfy the condition $\mathbf{char}(\F_q)>\deg \widehat{f}$. So if we directly use the Kronecker substitution,  it needs to make sure $\mathbf{char}(\F_q)>D^n$ which is a large characteristic. In order to improve the case, we use the sparse Kronecker substitution\cite{HG19}, the degree of the univariate polynomial after substitution is $O^\thicksim(n^2TD)$, so the characteristic of $\F_q$ only need to make sure $\mathbf{char}(\F_q)>O^\thicksim(n^2TD)$.

In this subsection, we use the sparse deterministic Kronecker substitutions to extended the univariate interpolation algorithm into a multivariate one.
Our  algorithm works as follows. 1:  Choose $O(nT\log D)$ primes $p_i$ of size $O^\thicksim(nT\log D)$ and substitutions $\mathbf{s}_i\in \Z^n$, where $\mathbf{s}_i=(1,D\ \modp\ p_i,\dots,D^{n-1}\ \modp\ p_i)$.
2: Find $i_0$ such that $\#f(x^{\mathbf{s}_{i_0}})\ \modp \ (x^{p_{i_0}}-1)$ has maximal number of terms.  3: Half of the terms of $f$ can be recovered from $f(x^{\mathbf{s}_{i_0}})$ and $f(x^{\mathbf{s}_{i_0}+p_{i_0}\mathbf{I}_k}),k=1,2,\dots,n$.

The degrees of $f(x^{\mathbf{s}_i}),i=1,\dots,N$ and $f(x^{\mathbf{s}_{i_0}+p_{i_0}\mathbf{I}_k}),k=1,\dots,n$ are bounded by $\max\{2np_iD|i=1,2,\dots,N\}$, where $N$ is a fix integer. Since $p_i$ is $O^\thicksim(nT\log D)$, the degree of these polynomials are bounded by $O^\thicksim(n^2TD)$. In order to recover polynomials $f(x^{\mathbf{s}_i})$ and $f(x^{\mathbf{s}_{i_0}+p_{i_0}\mathbf{I}_k})$ by Algorithm \ref{alg-9}, we need $\mathbf{char}(\F_q)>\max\{2np_iD|i=1,2,\dots,N\}$, which is $\mathbf{char}(\F_q)>O^\thicksim(n^2TD)$.

\vspace{11pt}
The following is the algorithm.

\begin{alg}[DMIPoly]\label{alg-10}
\end{alg}
{\noindent\bf Input:}
\begin{itemize}
\item A SLP $\S_f$ that computes $f\in\F_q[x_1,$ $\dots,x_n]$.

\item A terms bound $T\geq \# f$.

\item A partial degree bound $D> \max_{i=1}^n\deg_{x_i}(f)$.
\end{itemize}

{\noindent\bf Output:}
\begin{itemize}
\item If $\mathbf{char}(F_q)>2np_ND$,  return the exact form of $f$, where $p_N$ is defined in Step 2 and $2np_ND$ is  $O^\thicksim(n^2TD)$.
\item If $0<\mathbf{char}(\F_q)\leq 2np_ND$, return ``$\mathbf{char}(\F_q)$ is not large enough for this algorithm".

\end{itemize}

\begin{description}
\item[Step 1:] Let $N_1=\max\{1,\lceil n(T-1)\log_2 D\rceil\},N_2=\lceil nT\log_2 D\rceil,N=\max\{4N_1,N_1+N_2-1\}$, $T_1=T$, $h=0$.

\item[Step 2:] Find the first $N$ different primes $p_1,p_2,\dots,p_{N}$.

\item[Step 3:] For $i=1,2,\dots,N$, let $\mathbf{s}_i=(1,\modp(D,p_i),\dots,\modp(D^{n-1},p_i))$ and probe $f(x^{\mathbf{s}_i})_{p_i}$ from $\S_f$. Let $f_i=f(x^{\mathbf{s}_i})_{p_i}$.

\item[Step 4:] While $\max\{\#f_i|i=1,2,\dots,N\}\neq 0$ do

\begin{description}
\item[a:]Let $i_0$ be the smallest number such that $\#f_{i_0}=\max\{\#f_i|i=1,2,\dots,N\}$. Denote $\mathbf{s}_{i_0}=(s_1,\dots,s_n)$.

\item[b:] Denote $\Upsilon$ be the set of instructions  of $\S_f$  where  $x_i=x^{s_i},i=1,2,\dots,n$ are the inputs and $x$ is the new indeterminate. It is used to represent the univariate polynomial $f(x^{\mathbf{s}_{i_0}})$.
\item[c:] For $k=1,2,\dots,n$, denote $\Upsilon_k$ be the set of instructions  of $\S_f$  where  $x_i=x^{s_i},i=1,2,\dots,n,i\neq k$ and $x_k=x^{s_k+p}$ are the inputs and $x$ is the new indeterminate. They are used to represent the univariate polynomials $f(x^{\mathbf{s}_{i_0}+p_{i_0}\mathbf{I}_k}),k=1,2,\dots,n$.

\item[d:] $\{h^*_1,\dots,h^*_n\}=\mathbf{PolySubs}(h,\mathbf{s}_{i_0},p_{i_0})$.
\item[e:]
Let $F=\mathbf{DUIPoly}(\Upsilon,T,\|\mathbf{s}_{i_0}\|_\infty D)$ and $g=F-h(x^{\mathbf{s}_{i_0}})$.


\item[f:] For $k=1,2,\dots,n$, let $F_k=\mathbf{DUIPoly}(\Upsilon_k,T,\|\mathbf{s}_{i_0}\|_\infty D)$.
Let $g_k=F_k-h^*_k$.

\item[g:] Let $\hbox{TS}:=\mathbf{TSTerms}(f_{i_0}, g,g_1,g_2,\dots,g_n,p_{i_0},\mathbf{s}_{i_0},D)$.

\item[h:] Let $r=0$. For each $u\in \hbox{TS}$,
if
$$\#\{i\,|\, \#(f_i-u_{p_i})<\#(f_i),i=1,\dots,N_1+N_2-1 \}\ge N_2,$$
then $r:=r + u$.

\item[i:] Let $h=h+r$, $T_1=T_1-\#r$, $N_1=\max\{1,\lceil n(T_1-1)\log_2 D\rceil\},N_2=\lceil nT_1\log_2 D\rceil,N=\max\{4N_1,N_1+N_2-1\}$.

\item[j:] For $i=1,2,\dots,N$, let $f_i=f_i-r(x^{\mathbf{s}_i})_{p_i}$.

\end{description}

\item[Step 5:] Return $h$.

\end{description}

\begin{theorem}\label{the-13}
Algorithm \ref{alg-10} is correct. It needs $O^\thicksim(Ln^2T^2\log^2 D)$ field operations plus a similar number of bit operations.
\end{theorem}
\proof
For the correctness, we need to show that when $\mathbf{char}(\F_q)>2np_ND$, the algorithm returns $f$.
Since $\deg f(x^{\mathbf{s}_{i_0}})<npD$ and $\deg f(x^{\mathbf{s}_{i_0}+p\mathbf{I}_k})<2npD\leq 2np_ND$, $\mathbf{char}(\F_q)>\max\{\deg f(x^{\mathbf{s}_i}),i=1,\dots,N,\deg f(x^{\mathbf{s}_{i_0}+p_{i_0}\mathbf{I}_k}),k=1,\dots,n\}$. So by Theorem \ref{the-12}, in $\mathbf{e},\mathbf{f}$ of Step 4, $F=f(x^{\mathbf{s}_{i_0}})$ and $F_k=f(x^{\mathbf{s}_{i_0}+p\mathbf{I}_k})$.
In Step $\mathbf{g}$ of Step 4, by Theorem \ref{the-10} and Lemma \ref{lm-5}, at least half of terms of $f-h$ are in $\hbox{TS}$.
In $\mathbf{h}$ of Step 4, Theorem \ref{the-11} is used to select the elements of $M_{f-h}$ from $\hbox{TS}$.
Then the correctness of the algorithm is proved.

In Step 2, since the bit complexity of finding the first $N$ primes is $O(N\log^2N\log\log N)$ by \cite[p.500,Them.18.10]{GG99} and $N$ is $O^\thicksim(nT\log D)$, the bit complexity of Step 2 is $O^\thicksim(nT\log D)$.

In Step 3, compute each $\mathbf{s}_i$ needs $O(\log D+n\log p_i)$ bit operations. Since $p_i$ is $O^\thicksim(nT\log D)$ and $N$ is $O(nT\log D)$, it needs $O^\thicksim(nT\log^2 D+n^2T\log D)$ bit operations to compute all $\mathbf{s}_i$. For probing $f(x^{\mathbf{s}_i})_{p_i}$,  it probes $O(nT\log D)$ times. Since $p_i$ is  $O^\thicksim(nT\log D)$, by Lemma \ref{lm-1}, the cost of probes is $O^\thicksim(Ln^2T^2\log^2D)$ field operations and $O^\thicksim(n^2T\log D)$ bit operations.

Since every recursive of Step 4, at least half of terms in $f-h$ are found, it runs $O(\log T)$ times of Step 4. In $\mathbf{a}$, since $N$ is $O(nT\log D)$ and  $\#f_i\leq T$, it needs $O^\thicksim(nT^2\log D)$ bit operations to compute $\max\{\#f_i|i=1,2,\dots,N\}$.

In $\mathbf{d}$, by Lemma \ref{lm-7}, it needs $O(nT)$ arithmetic operations in $\F_q$ and $O^\thicksim(nT\log D)$ bit operations.

In $\mathbf{e},\mathbf{f}$, by Theorem \ref{the-12}, it needs $O^\thicksim(nLT^2\log^2(\|s_{i_0}D\|_\infty ))=O^\thicksim(LnT^2\log^2 D\log q)$ field operations and a similar number of bit operations.

In $\mathbf{g}$, by Lemma \ref{lm-6}, the complexity is $O(nT)$ field operations and $O^\thicksim(nT\log D)$ bit operations.

In $\mathbf{h}$, to compute all the $u_{p_j}$, we need $O^\thicksim(n(N_1+N_2)\#(\hbox{TS})\log (Dp_i))$ bit operations. The proof for rest of this step is similar to that of Step 8 of Algorithm \ref{alg-8}. The complexity is $O^\thicksim(n(N_1 + N_2)\#(\hbox{TS}) \log(Dp_i)+\#(\hbox{TS}) \log T (N_1+N_2)\log(Dp_i))$ bit operations and $O(\#(\hbox{TS})(N_1+N_2))$ field operations. Since $\#(\hbox{TS})\leq T, p_j=O^\thicksim(nT \log D)$, and $N_1 + N_2=O(nT \log D)$, it needs $O^\thicksim(n^2T^2\log^2 D)$ bit operations and $O^\thicksim(nT^2 \log D)$ field operations.

In $\mathbf{j}$, it needs $O(n\# r)$ operations in $\Z$ to obtain $r(x^{\mathbf{s}_i})_{p_i}$. Subtract $r(x^{\mathbf{s}_i})_{p_i}$ from $f_j$ needs $O(\# r\log T)$ operations in $\Z$ and $\#r$ arithmetic operation in $\F_q$. Since the height of the data is $O(\log (\|\mathbf{s}_i\|_\infty D))$ and we need update $N$ polynomials,  the complexity is $O((n\#r\log (\|\mathbf{s}_i\|_\infty D)+\#r\log T\log (\|\mathbf{s}_iD\|_\infty D)N)$ bit operations and $O(\#rN)$ field operations.
Since the sum of $\#r$ is $t$, it total costs $O^\thicksim(nT^2\log D)$ field operations and $O^\thicksim(n^2T^2\log^2 D)$ bit operations.
\qed

 \section{Conclusion}
In this paper, we consider sparse interpolation for a polynomial given by an SLP.
The main contributions are a Monte Carlo algorithm and a deterministic algorithms
which work over a finite field (for large characteristic). Our Monte Carlo has lower complexity than any existing algorithms if $q$ is $O((nTD)^{(1)})$.
Compare to our Monte Carlo algorithm, our deterministic algorithm probes more times and we also
give a criterion for checking whether a term belongs to a polynomial. The complexity of our deterministic algorithm is quadratic in the size of the sparse representation. We have showed that in finite field $\F_q$ with large characteristic, deterministic interpolation is as easy as Bl\"{a}ser et al. \cite{BHLV09} deterministic zero-testing for univariate polynomials represented by SLP.

\bibliographystyle{abbrv}
\bibliography{ref}

\end{document}